\documentclass[aps,pre,twocolumn,showpacs,groupeaddress,preprintnumbers,amsmath,amssymb,floatfix]{revtex4-1}

\usepackage{graphicx}
\usepackage{dcolumn}
\usepackage{bm}
\usepackage{hyperref}
\usepackage{color}

\begin{document}
\hspace{10in}
\title{Revival of oscillation from mean-field--induced death: Theory and experiment} 
\author{Debarati Ghosh${}^1$}, 
\author{Tanmoy Banerjee${}^{1}$}
\thanks{tbanerjee@phys.buruniv.ac.in}
\author{J\"urgen Kurths${}^{2,3,4,5}$}%
\affiliation{%
${}^1$Department of Physics, University of Burdwan, Burdwan 713 104, West Bengal, India.\\
${}^{2}$ Potsdam Institute for Climate Impact Research, Telegraphenberg, D-14415 Potsdam, Germany.\\
${}^3$ Institute of Physics, Humboldt University Berlin, D-12489 Berlin, Germany.\\
${}^4$  Institute for Complex Systems and Mathematical Biology, University of Aberdeen, Aberdeen AB24 3FX, UK.\\
${}^5$ Institute of Applied Physics of the Russian Academy of Sciences, 603950 Nizhny Novgorod, Russia.}%

\date{\today}

\begin{abstract}
The revival of oscillation and maintaining rhythmicity in a network of coupled oscillators offer an open challenge to researchers as the cessation of oscillation often leads to a fatal system degradation and an irrecoverable malfunctioning in many physical, biological and physiological systems.
Recently a general technique of restoration of rhythmicity in diffusively coupled networks of nonlinear oscillators  has been proposed in [Zou et al. Nature Commun. 6:7709, 2015], where it is shown that a proper feedback parameter that controls the rate of diffusion can effectively revive oscillation from an oscillation suppressed state. In this paper we show that the mean-field diffusive coupling, which can suppress oscillation even in a network of identical oscillators, can be modified in order to revoke the cessation of oscillation induced by it. Using a rigorous bifurcation analysis we show that, unlike other diffusive coupling schemes, here one has {\it two control parameters}, namely the {\it density of the mean-field} and the {\it feedback parameter} that can be controlled to revive oscillation from a death state. We demonstrate that an appropriate choice of density of the mean-field is capable of inducing rhythmicity even in the presence of complete diffusion, which is an unique feature of this mean-field coupling that is not available in other coupling schemes.  Finally, we report the {\it first} experimental observation of revival of oscillation from the mean-field--induced oscillation suppression state that supports our theoretical results.  
\end{abstract}


\maketitle 

\section{Introduction}
\label{sec:intro}
The suppression of oscillation and the  revival of oscillation are two opposite but interrelated important emergent phenomena in coupled oscillators.  In the suppression of oscillation, oscillators arrive at a common homogeneous steady state (HSS) or different branches of an inhomogeneous steady state (IHSS) under some proper parametric and coupling conditions. The former is called the amplitude death (AD) state \cite{adrev}, whereas the latter is denoted as the oscillation death (OD) state \cite{kosprep,*kosprl}. On the other hand, the revival of oscillation is a process by which the rhythmic behavior (or rhythmicity) of individual nodes in a network of coupled oscillators is restored from an AD/OD state without changing the intrinsic parameters associated with the individual nodes \cite{ryth2,*ryth3,*ryth5}.

In the oscillation quenched (suppressed) state the dynamic nature of individual coupled oscillators is lost. This has many potential applications: For example, the AD state is important to suppress unwanted oscillations that hinder a certain process, e.g., in laser systems, chattering in mechanical drilling process, etc \cite{adrev}. Similarly, the OD state has a significant role in understanding of many biological processes, e.g., synthetic genetic oscillator \cite{kosepl, qstr2}, cardiovascular phenomena \cite{Vargasepl}, cellular differentiation \cite{cell}, etc. On the other hand, the research in the topic of revival of oscillation is important because many physical, environmental, biological and social processes require {\it stable} and {\it robust} oscillations for their proper functioning: Examples include, El Ni\~{n}o/Southern Oscillation in ocean and atmosphere \cite{nino}, brain waves in neuroscience \cite{brain}, electric power generators \cite{kurths_pg},  cardiopulmonary sinus rhythm of pacemaker cells \cite{cardio}, etc. In these systems the suppression of oscillation may result in a fatal system breakdown or an irrecoverable physiological malfunction. Thus, it has to be ensured that, if these type of systems are trapped in an oscillation cessation state, that state has to be revoked in order to establish rhythmicity.  

A recent burst of publications reveal many aspects of AD and OD: In particular, identifying coupling schemes to induce them \cite{tanpre1,tanpre3,*scholl4,*kurthamp}, transition from AD to OD \cite{kosprep,*kosprl}, their experimental verifications \cite{tanpre2}, etc.  But only a few techniques have been reported to revoke AD/OD and induce rhythmicity in a network of oscillators \cite{ryth2,*ryth3,*ryth5,ryth1,*ryth4}. In Ref.\cite{ryth2,*ryth3,*ryth5} several variants of time delay techniques are discussed in order to revoke death states, whereas in Ref.\cite{ryth1,*ryth4}, network connections are chosen properly in order to revive oscillations. However, most of these techniques lack the generality to revive oscillations from a death state.   

Only recently a general technique to revive oscillation from the oscillation suppressed state (or death state) has been reported by \citet{kurthnat15}. The authors proposed a simple but effective way to revoke the death state and induce rhythmicity by introducing a simple feedback factor in the diffusive coupling. They showed that this technique is robust enough to induce  rhythmicity in a diffusively coupled network of nonlinear oscillators, such as, the Stuart-Landau oscillator, Brusselators model, chaotic Lorenz system, cell membrane model, etc. They further tested the effectiveness of their proposed technique in conjugate \cite{con} and dynamic coupling \cite{dyn}.

However, in the absence of parameter mismatch or coupling time-delay, simple diffusive coupling cannot induce AD in coupled oscillators. Therefore, for identical oscillators under diffusive coupling (without coupling delay) no AD is possible and thus the issue of revoking the AD state does not arise. Also, OD in diffusively coupled identical oscillators is always accompanied by a limit cycle, thus one needs to choose proper initial conditions to revoke that death state. Regarding the conjugate coupling, it is not always a general technique to induce death: For example, in a first-order intrinsic time-delay system no conjugate coupling is possible. Further, the dynamic coupling has its own pitfalls (e.g., its success strongly depends on the intrinsic properties of the oscillators under consideration), which has been discussed in detail in~\cite{karnatak_pit}. 

In this context the mean-field diffusive coupling is a general way to induce AD/OD even in networks of {\it identical} coupled oscillators: It works in any network of oscillators including chaotic first-order intrinsic time-delay systems \cite{tanchaosad}. Further, it has been shown in~\cite{tanpre1} that, unlike diffusive coupling, the OD state induced by the mean-field diffusive coupling is not accompanied by a limit cycle. Also, the mean-field diffusive coupling is the most natural coupling scheme that occurs in physics \cite{tanpre1,tanpre2,pathria}, biology \cite{bard,cell,qstr,qstr2}, ecology \cite{bandutta}, etc. Thus, for those systems that always need a robust limit cycle for their proper functioning, the mean-field diffusive coupling is a much stronger ``trap'' to induce death in comparison with the other coupling schemes. Therefore, it is important to study the process of revoking the oscillation suppression state induced by the mean-field diffusive coupling and revive oscillation from the death state.

Motivated by the above facts, in this paper we introduce a feedback factor in the {\it mean-field diffusive coupling} and examine its effect in a network of coupled oscillators. We show that the interplay of the {\it feedback factor} and the {\it density of mean-field} coupling can restore rhythmicity from a death state even in a network of identical coupled oscillators. Thus, unlike Ref.~\cite{kurthnat15}, here we have two control parameters that enable us to revoke the death state. Using rigorous eigenvalue and bifurcation analyses on coupled van der Pol and Stuart-Landau oscillators,  separately, we show that the region of the death state shrinks substantially in the parameter space depending upon those two control parameters. We also extend our results to a network consisting of a large number of mean-field coupled oscillators and show that the revival of rhythmicity works in the spatially extended systems, also. Further, for the first time, we report an experimental observation of the revival of oscillation from death states induced by the mean-field coupling that qualitatively supports our theoretical results. 

\section{Revival of oscillation with modified mean-field diffusive coupling: Theory}
\subsection{van der Pol oscillator}
\begin{figure*}[t!]
\includegraphics[width=.95\textwidth]{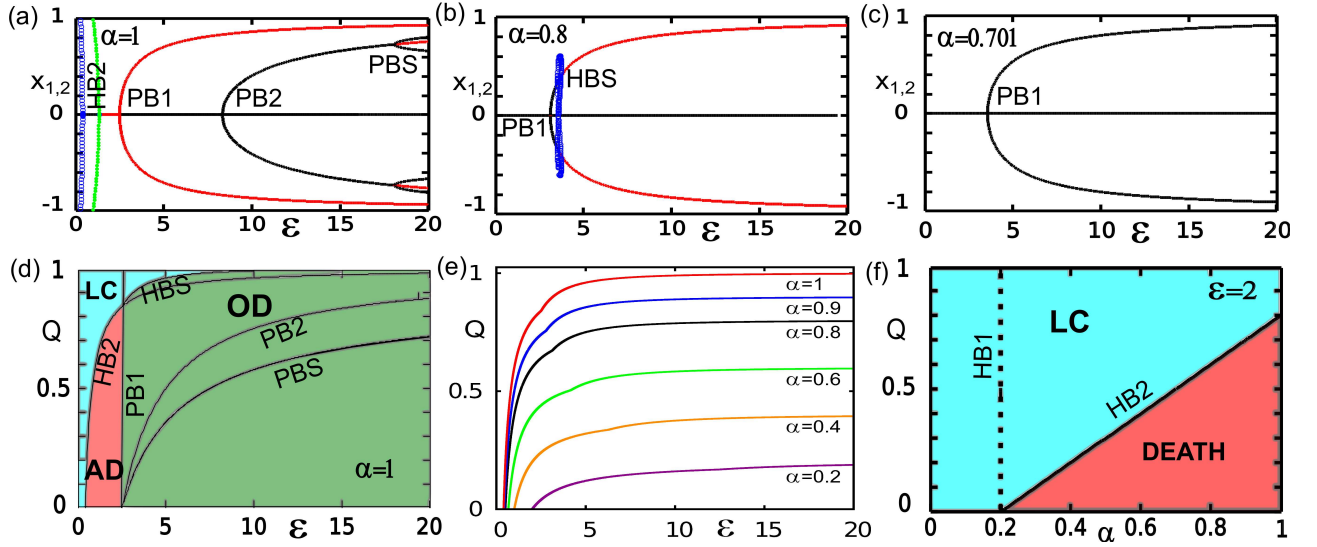}
\caption{\label{F:1} (Color online) Two ($N=2$) mean-field coupled van der Pol oscillators [Eqs.~\eqref{systemvdp}]: Bifurcation diagram with $\epsilon$ for (a) $\alpha=1$, (b) $\alpha=0.8$, (c) $\alpha=0.701$ at $Q=0.7$. Red (gray) line is for the stable fixed point, black lines are for unstable fixed points, green solid circle represents amplitude of stable limit cycle that emerges through Hopf bifurcation and blue open circle represents that of an unstable limit cycle. (d) Two parameter bifurcation diagram in the $\epsilon-Q$ space for $\alpha=1$ (e) Shrinking of death region in $\epsilon-Q$ space for decreasing $\alpha$. Area bellow each of the curves for a particular $\alpha$ represents the oscillation suppression region and above that shows the oscillating zone. (f) The variation of death region is shown in $\alpha-Q$ space using two parameter bifurcation  ($\epsilon=2$). (Other parameter: $a=0.4$).}
\end{figure*}
At first we consider a network of $N$ van der Pol  (VdP) oscillators interacting through a modified mean-field diffusive coupling; the mathematical model of the coupled system is given by
\begin{subequations}
\label{systemvdp}
\begin{align}
\label{x1}
\dot{x}_i &= y_i+\epsilon\left(Q\overline{X}-{\alpha}x_i\right),\\
\label{y1}
\dot{y}_i &= a_i(1-x{^2}_i)y_i-x_i.
\end{align}
\end{subequations}
Here $i=1\cdots N$ and $\overline{X}=\frac{1}{N}\sum_{i=1}^{N}x_i$ is the mean-field of the coupled system. The individual VdP oscillators  show a near sinusoidal oscillation for smaller $a_{i}$, and relaxation oscillation for larger $a_{i}$. The coupling strength is given by $\epsilon$;  $Q$ is called the mean-field density parameter that determines the {\it density of the mean-field} \cite{qstr,*qstr2,tanchaosad} ($ 0\le Q \le 1$); it actually provides an additional free parameter that controls the mean-field dynamics: $Q\rightarrow 0$ indicates the self-feedback case, whereas $Q\rightarrow 1$ represents the maximum mean-field density. The feedback term $\alpha$ controls the rate of diffusion ($0\le \alpha \le 1$): $\alpha =1$ represents the maximum feedback and the original mean-field diffusive coupling; $\alpha = 0$ represents the absence of a feedback and thus that of diffusion. Any values in between this limit can be treated as the intermediate diffusion rate and thus represent a modified mean-field diffusive coupling. The origin of $\alpha$ is well discussed in~\cite{kurthnat15} where it is speculated that it may arise in the context of cell cycle, neural network and synchronization engineering. 

As the limiting case we take two identical VdP oscillators: $a_{1,2}=a$. From Eq.\eqref{systemvdp} we can see that there are the following fixed points: the origin $(0, 0, 0, 0)$ and two more coupling dependent fixed points: (i) (${x}^\ast$, ${y}^\ast$, $-{x}^\ast$, $-{y}^\ast$) where ${x}^\ast =\frac{{y}^\ast}{{\epsilon}{\alpha}}$ and ${y}^\ast = \pm \sqrt {{\epsilon^2}{\alpha^2}-\frac{\epsilon{\alpha}}{a}}$. (ii) (${x}^\dagger$, ${y}^\dagger$, ${x}^\dagger$, ${y}^\dagger$) where ${x}^\dagger = \frac{{y}^\dagger}{\epsilon(\alpha - Q)}$ and ${y}^\dagger = \pm \sqrt {\epsilon^2(\alpha-Q)^2-\frac{\epsilon(\alpha-Q)}{a}}$.

The eigenvalues of the system at the origin are,
\begin{subequations}
\label{lambdavdp}
\begin{align}
\label{lambda1}
{\lambda}_{1,2} &= \frac{(a-\epsilon \alpha)\pm\sqrt{(a+\epsilon \alpha)^2-4}}{2},\\
\label{lambda3}
{\lambda}_{3,4} &= \frac{(a-\epsilon(\alpha -Q))\pm\sqrt{(a+\epsilon(\alpha -Q))^2-4}}{2}.
\end{align}
\end{subequations}
From the eigenvalue analysis we derive two pitchfork bifurcation (PB) points PB1 and PB2, which emerge at the following coupling strengths:
\begin{subequations}
\label{pb}
\begin{align}
\label{epsapb1}
{\epsilon}_{PB1} &= \frac{1}{a\alpha},\\
\label{epsapb2vdp}
{\epsilon}_{PB2} &= \frac{1}{a(\alpha -Q)}.
\end{align}
\end{subequations}
The IHSS, (${x}^\ast$, ${y}^\ast$, $-{x}^\ast$, $-{y}^\ast$), emerges at ${\epsilon}_{PB1}$ through a symmetry breaking pitchfork bifurcation. The other nontrivial fixed point (${x}^\dagger$, ${y}^\dagger$, ${x}^\dagger$, ${y}^\dagger$) comes into existence at ${\epsilon}_{PB2}$, which gives rise to an unique {\it nontrivial HSS}. Further, equating the real part of ${\lambda}_{1,2}$ and ${\lambda}_{3,4}$ to zero, we get two Hopf bifurcation points at 
\begin{subequations}
\label{hb12}
\begin{align}
\label{epsahb1}
{\epsilon}_{HB1} &= \frac{a}{\alpha},\\
\label{epsahb2}
{\epsilon}_{HB2} &= \frac{a}{(\alpha -Q)}.
\end{align}
\end{subequations}
From Eqs. \eqref{lambdavdp} and \eqref{hb12} we see that no Hopf bifurcation of trivial fixed point occurs for $a>1$; in that case, only pitchfork bifurcations exist.

The eigenvalues of the system at the nontrivial fixed point (${x}^j$, ${y}^j$, $J{x}^j$, $J{y}^j$), where $J=\pm 1$ and $j=\dagger~\mbox{or}~\ast$ are given by:
\begin{subequations}
\label{ntlambda}
\begin{align}
\label{ntlambda1-2}
{\lambda}_{1,2} &= \frac{{-b_1}^j \pm \sqrt{{{b_1}^j}^2-4{c_1}^j}}{2},\\
\label{ntlambda3-4}
{\lambda}_{3,4} &= \frac{{-b_2}^j \pm \sqrt{{{b_2}^j}^2-4{c_2}^j}}{2},
\end{align}
\end{subequations}
where, ${b_1}^j = \epsilon \alpha - a(1-{x^j}^2)$, ${c_1}^j = 1 + 2a{x^j}{y^j} - a\epsilon \alpha(1-{x^j}^2)$, ${b_2}^j = \epsilon(\alpha -Q) - a(1-{x^{j}}^2)$, ${c_2}^j = 1 + 2a{x^j}{y^j} - a\epsilon(\alpha -Q)(1-{x^j}^2)$.
Now, with increasing $Q$, ${\epsilon}_{HB2}$ moves towards ${\epsilon}_{PB1}$, and at a critical $Q$ value, say $Q^\ast$, HB2 collides with PB1:  $ Q^\ast = \alpha(1- a^2)$. For $Q>Q^\ast$, the IHSS becomes stable at $\epsilon_{HBS}$ through a subcritical Hopf bifurcation, where 
\begin{equation}
\label{epsahb3}
{\epsilon}_{HBS} = \frac{1}{\sqrt{\alpha(\alpha -Q)}}.
\end{equation}
This is derived from the eigenvalues of the system at the the nontrivial fixed point (${x}^\ast$, ${y}^\ast$, $-{x}^\ast$, $-{y}^\ast$). $\epsilon_{HBS}$ actually determines the direct transition from OD to a limit cycle, i.e. revival of oscillation. 

The second nontrivial fixed point $({x}^\dagger$, ${y}^\dagger$, ${x}^\dagger$, ${y}^\dagger)$ that was created at $\epsilon_{PB2}$ becomes stable through a subcritical pitchfork bifurcation at $\epsilon_{PBS}$: 
\begin{equation}\label{pbs}
{\epsilon}_{PBS} = \frac{(2\alpha -Q)}{2a(\alpha -Q)^2}.
\end{equation}
This is derived from the eigenvalues corresponding to $({x}^\dagger$, ${y}^\dagger$, ${x}^\dagger$, ${y}^\dagger)$. This nontrivial AD state can also be pushed back to a very large value of $\epsilon$ by choosing $\alpha \rightarrow Q$, and thus this AD state can also be revoked effectively.

The above eigenvalue analysis is supported by a numerical bifurcation analysis using XPPAUT \cite{xpp}.  Figure~\ref{F:1}(a-c) show the single parameter bifurcation diagram depending on $\epsilon$ for different $\alpha$ for an exemplary value $Q=0.7$ (throughout the numerical simulation we consider $a=0.4$). We observe that the oscillation cessation state (both AD and OD) moves towards right, i.e., a stronger coupling strength $\epsilon$ for a decreasing $\alpha$, and for $\alpha=0.701$ ($\rightarrow Q$) [Fig.~\ref{F:1} (c)] the death state moves much further from PB1 in the right direction. We also verify that at $\alpha=Q$ no death state occurs (not shown in the figure). We further explore the zone of oscillation cessation in a two parameter bifurcation diagram in the $\epsilon-Q$ space. Figure.~\ref{F:1} (d) shows the bifurcation curves for $\alpha=1$. The HB2 curve determines the transition from oscillation to AD for $Q<Q^\ast$; beyond this limit the HBS curve determines the zone of oscillation and thus that of the death region. Figure.~\ref{F:1} (e)  shows that the {\it death region shrinks with decreasing $\alpha$} confirming our theoretical analysis. Finally, we plot the phase diagram in $Q-\alpha$ parameter space at $\epsilon=2$ [Fig.~\ref{F:1} (f)]: we find that a {\it higher} value of $Q$ {\it or} a {\it lower} value of $\alpha$ support oscillations. Interestingly, even for $\alpha=1$ (i.e., a complete mean-field diffusion), one can revive rhythmicity by simply increasing the value of $Q$; thus, this coupling scheme offers two control parameters to revive oscillations.

Finally, we summarize the observations and discuss the following important points:
(i)  HB2 is the inverse Hopf bifurcation point where an AD state is revoked and gives rise to a {\it stable} limit cycle. This point (or curve in a two parameter space) determines the revival of oscillation below a critical value $Q^\ast$, which is determined by $\alpha$. From Eq.~\eqref{epsahb3} we see that by choosing $\alpha$ closer to $Q$ ($\alpha>Q$) the death zone shrinks substantially. Thus, to revoke a death state one has to choose $\alpha \rightarrow Q$, and $\alpha=Q$ {\it ensures} that there will be no death state even in the stronger coupling strength (whatever strong it may be).
(ii) Even if complete diffusion is present, i.e., $\alpha=1$, one can still achieve the revival of oscillation by choosing $Q\rightarrow1$. This is an unique feature of the mean-field diffusive coupling, which is absent in other coupling schemes. 
\begin{figure}
\includegraphics[width=.48\textwidth]{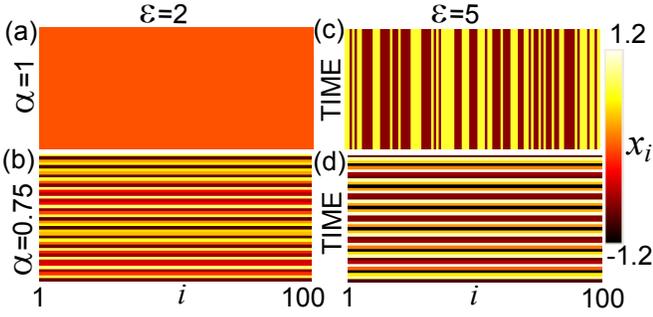}
\caption{\label{F:2} (Color online) Network of mean-field coupled van der Pol oscillators [Eqs.~\eqref{systemvdp}] with $Q=0.7$, $a=0.4$, $N=100$: Spatiotemporal plots showing  (a) and (b) AD and its revival, respectively, with decreasing $\alpha$ at $\epsilon=2$. (c) and d) show  the OD and its revival, respectively, with a decreasing $\alpha$ at $\epsilon=5$. The upper rows [i.e., (a) and (c)] have $\alpha=1$ and the lower rows [i.e., (b) and (d)] have $\alpha=0.75$. The first $t=5000$ time is excluded and then $x_i$s for the next $t=100$ are plotted.}
\end{figure}

To show that our analysis of two coupled oscillators are valid for a larger network also, we consider the more general case of $N=100$ mean-field coupled identical van der Pol oscillators of Eqs.~\eqref{systemvdp} ($a=0.4$). Figure.~\ref{F:2}(a) shows the spatiotemporal plot of the network in the global AD regime at $\alpha=1$, $Q=0.7$ and $\epsilon=2$; here all the nodes arrive at the zero fixed point. The global AD state is revoked and rhythmicity is restored in the network by decreasing the value of $\alpha$; as shown in Fig.~\ref{F:2}(b) for an exemplary value $\alpha=0.75$. Equivalently, for $\epsilon=5$ and $\alpha=1$ ($Q=0.7$ as before) we get an OD state in the network [Fig.~\ref{F:2}(c)]. It can be seen that the nodes populate the upper and lower branches [shown with yellow (light gray) and brown (dark gray) colors, respectively] of OD in a random manner and generate a multi-cluster OD state. Oscillation in this network is revived by decreasing $\alpha$; Fig.~\ref{F:2}(d) shows this for $\alpha=0.75$. Note that the values for which AD, OD and oscillations are obtained agree with that for the $N=2$ case shown in Fig.~\ref{F:1}.

\subsection{Stuart-Landau oscillator}
\label{sslo}
Next, we consider $N$ Stuart-Landau oscillators interacting through a modified mean-field diffusive coupling in their real part; the mathematical model of the coupled system is given by
\begin{equation}\label{ls} 
\dot{Z_i}=(1+i\omega_i-|Z_i|^{2})Z_i+\epsilon\bigg(Q\overline{Z}-{\alpha}Re(Z_i)\bigg),
\end{equation}
with $i=1\cdots N$; $\overline{Z}=\frac{1}{N}\sum_{i=1}^{N}Re(Z_i)$ is the mean-field of the coupled system, $Z_i=x_i+jy_i$. The individual Stuart-Landau oscillators are of unit amplitude and having eigenfrequency $\omega_i$. As the limiting case we take $N=2$, and rewrite Eq.~\eqref{ls} in the Cartesian coordinates:
\begin{subequations}
\label{system}
\begin{align}
\label{x1}
\dot{x}_{i} &= P_{i}x_{i}-\omega_{i}y_{i}+\epsilon[Q\overline{X}-{\alpha}x_{i}],\\
\label{y1}
\dot{y}_{i} &= \omega_{i}x_{i}+P_{i}y_{i}.
\end{align}
\end{subequations}
Here $P_i = 1-{x_i}^2-{y_i}^2$ $(i = 1,2)$, $\overline{X} = \frac{x_1+x_2}{2}$. We set the oscillators to be identical, i.e., $\omega_{1,2}=\omega$. From Eq.~\eqref{system} it is clear that the system has the following fixed points: the trivial fixed point is the origin $(0, 0, 0, 0)$, and additionally two more coupling dependent nontrivial fixed points: (i) (${x}^\ast$, ${y}^\ast$, $-{x}^\ast$, $-{y}^\ast$) where ${x}^\ast = -\frac{\omega {y}^\ast}{{\omega}^2 + \epsilon \alpha {{y}^\ast}^2}$ and ${y}^\ast =\pm \sqrt {\frac{(\epsilon \alpha - 2{\omega}^2) + \sqrt{{\epsilon}^2{\alpha}^2 - 4{\omega}^2}}{2\epsilon \alpha}}$. (ii) (${x}^\dagger$, ${y}^\dagger$, ${x}^\dagger$, ${y}^\dagger$) where ${x}^\dagger = - \frac{\omega {y}^\dagger}{\epsilon(\alpha - Q){{y}^\dagger}^2 + {\omega}^2}$ and ${y}^\dagger =\pm \sqrt{\frac{\epsilon(\alpha - Q) - 2 {\omega}^2 + \sqrt{{{\epsilon}^2 (\alpha -Q)}^2 - 4{\omega}^2}}{2\epsilon(\alpha - Q)}}$.
\begin{figure}
\includegraphics[width=.48\textwidth]{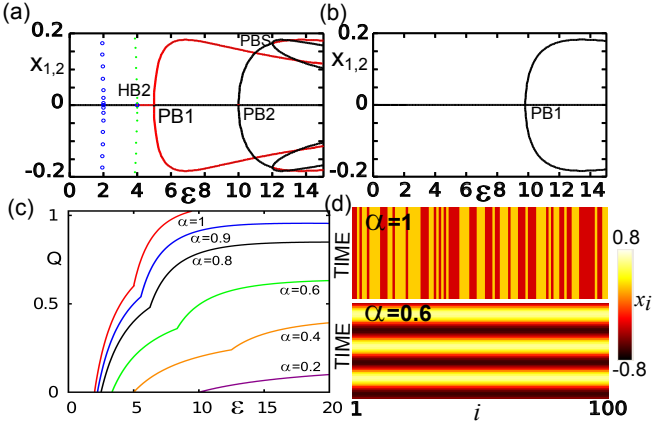}
\caption{\label{F:3} (Color online) Stuart-Landau oscillators, $\omega=2$: Bifurcation diagram with $\epsilon$ (a) $\alpha=1$ ($Q=0.5$), (b) $\alpha=0.5$ ($Q=0.5$). (c) Two parameter bifurcation diagram in $\epsilon-Q$ space for different $\alpha$. (d) Spatiotemporal plots of $N=100$ Stuart-Landau oscillators  showing OD at $\alpha=1$ (upper panel) and the revival of oscillation at $\alpha=0.6$ (lower panel): $Q=0.5$ and $\epsilon=6$. The time scale is same as in Fig.~\ref{F:2}.}
\end{figure}

The four eigenvalues of the system at the trivial fixed point $(0,0,0,0)$ are,
\begin{subequations}
\label{lambda}
\begin{align}
\label{lambda1}
{\lambda}_{1,2} &= 1 - \left[\frac{\epsilon(\alpha -Q)\pm\sqrt{{\epsilon}^2(\alpha -Q)^2-4{\omega}^2}}{2}\right],\\
\label{lambda3}
{\lambda}_{3,4} &= 1- \left[\frac{\epsilon \alpha \pm \sqrt{(\epsilon \alpha)^2-4{\omega}^2}}{2}\right].
\end{align}
\end{subequations}
Through an eigenvalue analysis and also a close inspection of the nontrivial fixed points reveal that two pitchfork bifurcations (PB) occur at:
\begin{subequations}
\label{pb}
\begin{align}
\label{epsapb1}
{\epsilon}_{PB1} &= \frac{1+{\omega}^2}{\alpha},\\
\label{epsapb2}
{\epsilon}_{PB2} &= \frac{1+{\omega}^2}{\alpha -Q}.
\end{align}
\end{subequations}
A symmetry breaking pitchfork bifurcation gives birth to the IHSS  (${x}^\ast$, ${y}^\ast$, $-{x}^\ast$, $-{y}^\ast$) at ${\epsilon}_{PB1}$. The second nontrivial fixed point (${x}^\dagger$, ${y}^\dagger$, ${x}^\dagger$, ${y}^\dagger$) emerges at PB2 the stabilization of which leads to a {\it nontrivial} AD state that coexists with OD.  

Next, we get the Hopf bifurcation point by equating the real part of ${\lambda}_{3,4}$ and ${\lambda}_{1,2}$ to zero, 
\begin{subequations}
\label{epsahb}
\begin{align}
{\epsilon}_{HB1} &= \frac{2}{\alpha},\\
{\epsilon}_{HB2} &= \frac{2}{\alpha -Q},
\end{align}
\end{subequations}
From Eqs. \eqref{lambda} and \eqref{epsahb} it is clear that for $\omega\le1$ no Hopf bifurcations (of trivial fixed point) occur, only pitchfork bifurcations govern the dynamics in that case. From Eq.~(\ref{epsahb}) it is noticed that for a fixed $\alpha$,  ${\epsilon}_{HB1}$ is constant, but ${\epsilon}_{HB2}$ depends only upon $Q$ (but is independent of $\omega$, where $\omega>1$). Now with increasing $Q$, HB2 moves towards PB1, and at a critical $Q$, say $Q^\ast$, HB2 collides with PB1: $Q^\ast = \frac{\alpha({\omega}^2-1)}{{\omega}^2+1}$. 

The eigenvalues of the system at the nontrivial fixed point (${x}^j$, ${y}^j$, $J{x}^j$, $J{y}^j$), where $J=\pm 1$ and $j=\dagger~\mbox{or}~\ast$ are given by:
\begin{subequations}
\label{ntlambda}
\begin{align}
\label{ntlambda1-2}
{\lambda}_{1,2} &= \frac{{-b_1}^j \pm \sqrt{{{b_1}^j}^2-4{c_1}^j}}{2},\\
\label{ntlambda3-4}
{\lambda}_{3,4} &= \frac{{-b_2}^j \pm \sqrt{{{b_2}^j}^2-4{c_2}^j}}{2}.
\end{align}
\end{subequations}
Where, ${b_1}^j = 4({x^j}^2+{y^j}^2)-2+\epsilon(\alpha -Q)$, ${c_1}^j = {\omega}^2 - 4{x^j}^2{y^j}^2 +(1-{x^j}^2-3{y^j}^2)(1-\epsilon \alpha +\epsilon Q - 3{x^j}^2-{y^j}^2)$, ${b_2}^j = 4({x^j}^2+{y^j}^2)-2+\epsilon \alpha$, ${c_2}^j = {\omega}^2 - 4{x^j}^2{y^j}^2 +(1-{x^j}^2-3{y^j}^2)(1-\epsilon \alpha - 3{x^j}^2-{y^j}^2)$. We derive the loci of the HBS curve as
\begin{equation}
\label{hbs-slo}
{\epsilon}_{HBS} = \frac{-2(Q+\alpha)+4\sqrt{{\alpha}^2+{\omega}^2(\alpha -Q)(3\alpha +Q)}}{(\alpha -Q)(3\alpha +Q)}.
\end{equation}
Bifurcation diagrams of Fig.~\ref{F:3}(a) ($\alpha=1$) and \ref{F:3}(b) ($\alpha=0.51$) show that with decreasing $\alpha$ the death region moves towards a larger coupling strength $\epsilon$. Figure~\ref{F:3}(c) shows this in the $\epsilon-Q$ space; one can see that a decreasing $\alpha$ shrinks the region of death and thus broadens the area of the oscillation state in the parameter space. Finally, we consider a network of $N=100$ identical Stuart-Landau oscillators with the same coupling scheme [Eq.~\eqref{ls}]. We find that the revival of oscillation works here as well. Figure~\ref{F:3}(d) demonstrates this for an exemplary value of $\epsilon=6$ and $Q=0.5$:  The upper panel shows the OD state in the network for $\alpha=1$, whereas from the lower panel we see that rhytmicity is restored in the network by reducing the value of $\alpha$ to $\alpha=0.6$. 

\section{Experimental observation of the revival of oscillation}
\label{expt}
\begin{figure}
\includegraphics[width=.42\textwidth]{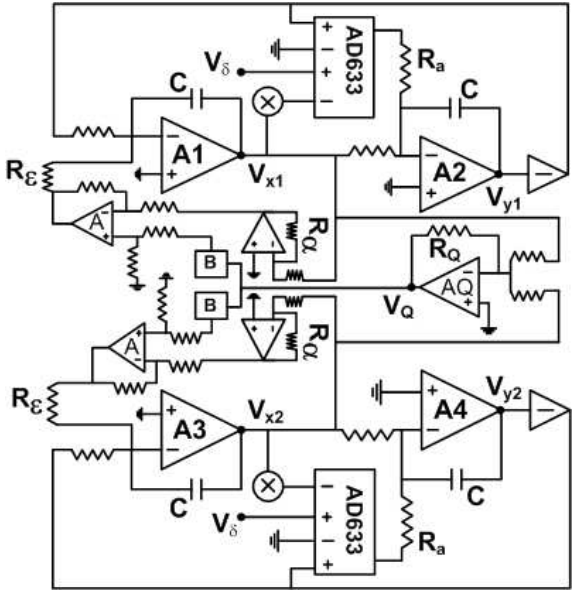}
\caption{\label{ckt} (Color online) Experimental circuit diagram of the modified mean-field coupled VdP oscillators. A, A1-A4, and AQ are realized with TL074 op-amps. All the unlabeled resistors have value $R=10$~k$\Omega$. C=10 nF, $R_a=250\Omega$, $V_\delta=0.1$ v. Box denoted by ``B" are op-amp based buffers; inverters are realized with the unity gain inverting op-amps. $\otimes$ sign indicates squarer using AD633.}
\end{figure}
Next, we implement the coupled system of van der Pol oscillators given by Eq.~\eqref{systemvdp} in an electronic circuit (Figure~\ref{ckt}). We use TL074 (quad JFET) op-amps, and AD633 analog multiplier ICs. A $\pm15$ v power supply is used; resistors (capacitors) have $\pm5\%$ ($\pm1\%$) tolerance. The unlabeled resistors have the value $R=10$~k$\Omega$. The op-amp AQ is used to generate the mean-field: $V_Q=-\frac{2R_Q}{R}\sum_{j=1}^{2}\frac{V_{xj}}{2}$, which is subtracted by $\frac{R_{\alpha}}{R}V_{x1,2}$ using op-amps denoted by A. One can see that $R_\epsilon$ determines the coupling strength, $R_Q$ determines the mean-field density and $R_{\alpha}$ controls the feedback parameter $\alpha$. The voltage equation of the circuit can be written as:
\begin{subequations}\label{ckteqn}
\begin{align}
CR\frac{d{V}_{xi}}{dt}&=V_{yi}+\frac{R}{R_\epsilon}\left[\frac{2R_Q}{R}\sum_{j=1}^{2}\frac{V_{xj}}{2}-\frac{R_{\alpha}}{R}V_{xi}\right],\\
CR\frac{d{V}_{yi}}{dt}&=\frac{R}{R_a}\left(V_\delta-\frac{V_{xi}^2}{10}\right)\frac{V_{yi}}{10}-V_{xi}.
\end{align}
\end{subequations}
Here $i=1,2$. Eqs.~\eqref{ckteqn} is normalized with respect to $CR$, and thus now becomes equivalent to Eq.~\eqref{system} for the following normalized parameters: $\dot{u}=\frac{du}{d\tau}$, $\tau=t/RC$, $\epsilon=\frac{R}{R_\epsilon}$, $Q=\frac{2R_Q}{R}$, $\alpha=\frac{R_{\alpha}}{R}$, $a=\frac{R}{100R_a}$, $10V_\delta=1$, $x_i=\frac{V_{xi}}{V_{sat}}$, and $y_i=\frac{V_{yi}}{V_{sat}}$. $V_{sat}$ is the saturation voltage of the op-amp. In the experiment we take $V_\delta=0.1$ v, and $C=10$ nF; we choose $a=0.4$ by taking $R_a=250$~$\Omega$ [using a precision potentiometer (POT)]. 

\begin{figure}
\includegraphics[width=.48\textwidth]{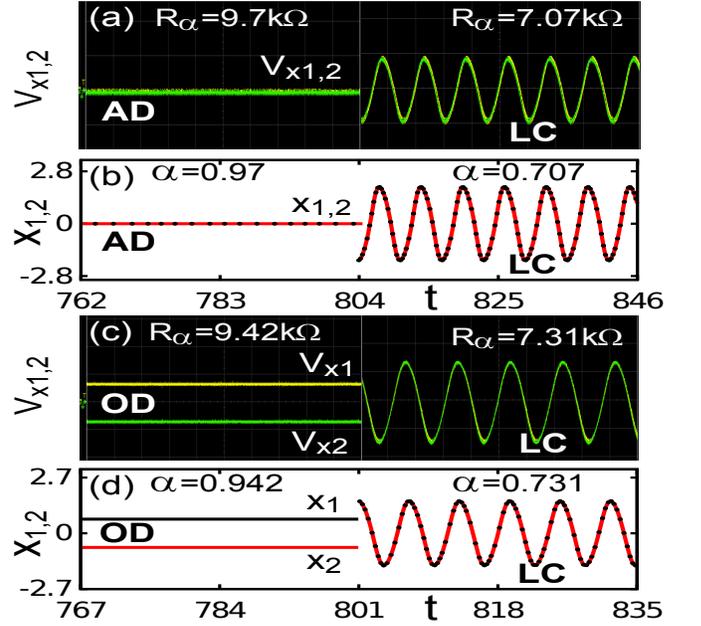}
\caption{\label{F:5} (Color online) (a, c) Experimental real time traces of $V_{x1}$ and $V_{x2}$ along with the (b, d) numerical time series plots of $x_1$ and $x_2$. [(a) and (b)] With $R_{\epsilon}=5$~k$\Omega$ (i.e. $\epsilon = 2$) a decreasing $R_{\alpha}$ ($\alpha$) restores oscillation (LC) from AD: AD at $R_{\alpha} = 9.7$~k$\Omega$ ($\alpha = 0.97$), LC at $R_{\alpha} = 7.07$~k$\Omega$ ($\alpha = 0.707$). [(c) and (d)] With $R_{\epsilon}=2$~k$\Omega$ (i.e. $\epsilon = 5$) a decreasing $R_{\alpha}$ ($\alpha$) restores oscillation (LC) from OD: OD at $R_{\alpha} = 9.42$~k$\Omega$ ($\alpha = 0.942$), LC at $R_{\alpha} = 7.31$~k$\Omega$ ($\alpha = 0.731$). Others parameters are $R_{Q} = 3.5$~k$\Omega$ ($Q=0.7$) and $R_{a} = 250$~$\Omega$ ($a=0.4$). $y$ axis: (a) 200 mv/div (c) 100 mv/div; $x$ axis: 380 $\mu$s/div.}
\end{figure}

We experimentally observe the revival of oscillation by revoking the oscillation cessation states (AD and OD) with varying $\alpha$ (i.e., $R_\alpha$). At first we consider the case of the AD state: For that we choose $\epsilon=2$ and $Q=0.7$ (by setting $R_{\epsilon}=5$~k$\Omega$ and $R_{Q} = 3.5$~k$\Omega$, respectively) and decrease $\alpha$ from $\alpha=1$ to a lower value. Figure~\ref{F:5}(a) shows experimental snapshots of the AD state in $V_{x1}$ and $V_{x2}$ at $R_{\alpha} = 9.7$~k$\Omega$ (i.e., $\alpha = 0.97$) [using DSO, Agilent make, DSO-X 2024A, 200 MHz, 2 Gs/s]; in the same figure we show the revival of oscillation from the AD state at an exemplary value $R_{\alpha} = 7.07$~k$\Omega$ ($\alpha = 0.707$). Figure~\ref{F:5}(b) gives the numerical time series at the corresponding $\alpha$ values (Fourth-order Runge-Kutta method with step-size of $0.01$). The numerical results are in good agreement with the experimental observations: Also, the dynamical behaviors at these parameter values are in accordance with Fig.~\ref{F:1}(f). 

Next, we choose an OD state for $\epsilon=5$ and $Q=0.7$ (i.e., $R_{\epsilon}=2$~k$\Omega$ and $R_{Q} = 3.5$~k$\Omega$, respectively): Experimental snapshots of the OD state at $R_{\alpha} = 9.42$~k$\Omega$ ($\alpha = 0.942$) and the rhythmicity at $R_{\alpha} = 7.31$~k$\Omega$ ($\alpha = 0.731$) are shown in Fig.~\ref{F:5}(c). The corresponding numerical result is given in Fig.~\ref{F:5}(d). We see that the experimental and numerical results are in good agreement and in accordance with Fig.~\ref{F:1}(e). Significantly, despite the presence of inherent parameter fluctuations and noise, which is inevitable in an experiment, it is important that in both the above cases theory and experiment are in good qualitative agreement, which proves the robustness of the coupling scheme in restoring rhythmicity in coupled oscillators.  
\section{CONCLUSION}
\label{sec:con}
We have investigated the effect of a feedback parameter, which controls the diffusion rate, on a network of nonlinear oscillators coupled under mean-field diffusion. We have shown that, unlike other coupling schemes, here two control parameters exist, namely the density of mean-field diffusion and the feedback parameter. The interplay of these two parameters can revive rhythmicity from any oscillation cessation state: In fact by controlling the feedback parameter closer to the density of the mean-field one can shrink the region of the oscillation cessation state to a very narrow zone in parameter space. More interestingly, even in the presence of complete diffusion (i.e., in the absence of feedback parameter), the density of the mean-field alone can induce rhythmicity from a death state. Thus, it offers a very robust oscillation revival mechanism. We have extended our study to a network consists of large number of nodes and shown that the oscillation cessation states can be revoked in that case too. Finally, we have supported our theoretical results by an experiment with electronic van der Pol oscillators and  for the first time observed the revival of oscillation from the mean-field-diffusion--induced death state. 

Since both the density of the mean-field and the feedback parameter have strong connections with many real biological networks including cell cycle, neural network, synthetic genetic oscillators, etc., we believe that the present study will broaden our understanding of those systems and subsequently this study will shed light on the control of oscillation suppression mechanism in several biological and engineering systems. 


\begin{acknowledgments}
T. B. acknowledges the financial support from SERB, Department of Science and Technology (DST), India [Project Grant No.: SB/FTP/PS-005/2013]. D. G. acknowledges DST, India, for providing support through the INSPIRE fellowship. J. K. acknowledges Government of the Russian Federation (Agreement No. 14.Z50.31.0033 with Institute of Applied Physics RAS).
\end{acknowledgments}



\begin{thebibliography}{31}%
\makeatletter
\providecommand \@ifxundefined [1]{%
 \@ifx{#1\undefined}
}%
\providecommand \@ifnum [1]{%
 \ifnum #1\expandafter \@firstoftwo
 \else \expandafter \@secondoftwo
 \fi
}%
\providecommand \@ifx [1]{%
 \ifx #1\expandafter \@firstoftwo
 \else \expandafter \@secondoftwo
 \fi
}%
\providecommand \natexlab [1]{#1}%
\providecommand \enquote  [1]{``#1''}%
\providecommand \bibnamefont  [1]{#1}%
\providecommand \bibfnamefont [1]{#1}%
\providecommand \citenamefont [1]{#1}%
\providecommand \href@noop [0]{\@secondoftwo}%
\providecommand \href [0]{\begingroup \@sanitize@url \@href}%
\providecommand \@href[1]{\@@startlink{#1}\@@href}%
\providecommand \@@href[1]{\endgroup#1\@@endlink}%
\providecommand \@sanitize@url [0]{\catcode `\\12\catcode `\$12\catcode
  `\&12\catcode `\#12\catcode `\^12\catcode `\_12\catcode `\%12\relax}%
\providecommand \@@startlink[1]{}%
\providecommand \@@endlink[0]{}%
\providecommand \url  [0]{\begingroup\@sanitize@url \@url }%
\providecommand \@url [1]{\endgroup\@href {#1}{\urlprefix }}%
\providecommand \urlprefix  [0]{URL }%
\providecommand \Eprint [0]{\href }%
\providecommand \doibase [0]{http://dx.doi.org/}%
\providecommand \selectlanguage [0]{\@gobble}%
\providecommand \bibinfo  [0]{\@secondoftwo}%
\providecommand \bibfield  [0]{\@secondoftwo}%
\providecommand \translation [1]{[#1]}%
\providecommand \BibitemOpen [0]{}%
\providecommand \bibitemStop [0]{}%
\providecommand \bibitemNoStop [0]{.\EOS\space}%
\providecommand \EOS [0]{\spacefactor3000\relax}%
\providecommand \BibitemShut  [1]{\csname bibitem#1\endcsname}%
\let\auto@bib@innerbib\@empty
\bibitem [{\citenamefont {Saxena}\ \emph {et~al.}(2012)\citenamefont {Saxena},
  \citenamefont {Prasad},\ and\ \citenamefont {Ramaswamy}}]{adrev}%
  \BibitemOpen
  \bibfield  {author} {\bibinfo {author} {\bibfnamefont {G.}~\bibnamefont
  {Saxena}}, \bibinfo {author} {\bibfnamefont {A.}~\bibnamefont {Prasad}}, \
  and\ \bibinfo {author} {\bibfnamefont {R.}~\bibnamefont {Ramaswamy}},\
  }\href@noop {} {\bibfield  {journal} {\bibinfo  {journal} {Phys. Rep.}\
  }\textbf {\bibinfo {volume} {521}},\ \bibinfo {pages} {205} (\bibinfo {year}
  {2012})}\BibitemShut {NoStop}%
\bibitem [{\citenamefont {Koseska}\ \emph
  {et~al.}(2013{\natexlab{a}})\citenamefont {Koseska}, \citenamefont {Volkov},\
  and\ \citenamefont {Kurths}}]{kosprep}%
  \BibitemOpen
  \bibfield  {author} {\bibinfo {author} {\bibfnamefont {A.}~\bibnamefont
  {Koseska}}, \bibinfo {author} {\bibfnamefont {E.}~\bibnamefont {Volkov}}, \
  and\ \bibinfo {author} {\bibfnamefont {J.}~\bibnamefont {Kurths}},\
  }\href@noop {} {\bibfield  {journal} {\bibinfo  {journal} {Phys. Rep.}\
  }\textbf {\bibinfo {volume} {531}},\ \bibinfo {pages} {173} (\bibinfo {year}
  {2013}{\natexlab{a}})}\BibitemShut {NoStop}%
\bibitem [{\citenamefont {Koseska}\ \emph
  {et~al.}(2013{\natexlab{b}})\citenamefont {Koseska}, \citenamefont {Volkov},\
  and\ \citenamefont {Kurths}}]{kosprl}%
  \BibitemOpen
  \bibfield  {author} {\bibinfo {author} {\bibfnamefont {A.}~\bibnamefont
  {Koseska}}, \bibinfo {author} {\bibfnamefont {E.}~\bibnamefont {Volkov}}, \
  and\ \bibinfo {author} {\bibfnamefont {J.}~\bibnamefont {Kurths}},\
  }\href@noop {} {\bibfield  {journal} {\bibinfo  {journal} {Phy. Rev. Lett}\
  }\textbf {\bibinfo {volume} {111}},\ \bibinfo {pages} {024103} (\bibinfo
  {year} {2013}{\natexlab{b}})}\BibitemShut {NoStop}%
\bibitem [{\citenamefont {Zou}\ \emph {et~al.}(2013)\citenamefont {Zou},
  \citenamefont {Senthilkumar}, \citenamefont {Zhan},\ and\ \citenamefont
  {Kurths}}]{ryth2}%
  \BibitemOpen
  \bibfield  {author} {\bibinfo {author} {\bibfnamefont {W.}~\bibnamefont
  {Zou}}, \bibinfo {author} {\bibfnamefont {D.~V.}\ \bibnamefont
  {Senthilkumar}}, \bibinfo {author} {\bibfnamefont {M.}~\bibnamefont {Zhan}},
  \ and\ \bibinfo {author} {\bibfnamefont {J.}~\bibnamefont {Kurths}},\
  }\href@noop {} {\bibfield  {journal} {\bibinfo  {journal} {Phys. Rev. Lett.}\
  }\textbf {\bibinfo {volume} {111}},\ \bibinfo {pages} {014101} (\bibinfo
  {year} {2013})}\BibitemShut {NoStop}%
\bibitem [{\citenamefont {Konishi}(2005)}]{ryth3}%
  \BibitemOpen
  \bibfield  {author} {\bibinfo {author} {\bibfnamefont {K.}~\bibnamefont
  {Konishi}},\ }\href@noop {} {\bibfield  {journal} {\bibinfo  {journal} {Phys.
  Lett. A}\ }\textbf {\bibinfo {volume} {341}},\ \bibinfo {pages} {401}
  (\bibinfo {year} {2005})}\BibitemShut {NoStop}%
\bibitem [{\citenamefont {Zou}\ \emph {et~al.}(2010)\citenamefont {Zou},
  \citenamefont {Yao},\ and\ \citenamefont {Zhan}}]{ryth5}%
  \BibitemOpen
  \bibfield  {author} {\bibinfo {author} {\bibfnamefont {W.}~\bibnamefont
  {Zou}}, \bibinfo {author} {\bibfnamefont {C.}~\bibnamefont {Yao}}, \ and\
  \bibinfo {author} {\bibfnamefont {M.}~\bibnamefont {Zhan}},\ }\href@noop {}
  {\bibfield  {journal} {\bibinfo  {journal} {Phys. Rev. E}\ }\textbf {\bibinfo
  {volume} {82}},\ \bibinfo {pages} {056203} (\bibinfo {year}
  {2010})}\BibitemShut {NoStop}%
\bibitem [{\citenamefont {Koseska}\ \emph {et~al.}(2009)\citenamefont
  {Koseska}, \citenamefont {Volkov},\ and\ \citenamefont {Kurths}}]{kosepl}%
  \BibitemOpen
  \bibfield  {author} {\bibinfo {author} {\bibfnamefont {A.}~\bibnamefont
  {Koseska}}, \bibinfo {author} {\bibfnamefont {E.}~\bibnamefont {Volkov}}, \
  and\ \bibinfo {author} {\bibfnamefont {J.}~\bibnamefont {Kurths}},\
  }\href@noop {} {\bibfield  {journal} {\bibinfo  {journal} {Euro. Phys.
  Lett.}\ }\textbf {\bibinfo {volume} {85}},\ \bibinfo {pages} {28002}
  (\bibinfo {year} {2009})}\BibitemShut {NoStop}%
\bibitem [{\citenamefont {Ullner}\ \emph {et~al.}(2007)\citenamefont {Ullner},
  \citenamefont {Zaikin}, \citenamefont {Volkov},\ and\ \citenamefont
  {Garc{\'{i}}a-Ojalvo}}]{qstr2}%
  \BibitemOpen
  \bibfield  {author} {\bibinfo {author} {\bibfnamefont {E.}~\bibnamefont
  {Ullner}}, \bibinfo {author} {\bibfnamefont {A.}~\bibnamefont {Zaikin}},
  \bibinfo {author} {\bibfnamefont {E.~I.}\ \bibnamefont {Volkov}}, \ and\
  \bibinfo {author} {\bibfnamefont {J.}~\bibnamefont {Garc{\'{i}}a-Ojalvo}},\
  }\href@noop {} {\bibfield  {journal} {\bibinfo  {journal} {Phy. Rev. Lett.}\
  }\textbf {\bibinfo {volume} {99}},\ \bibinfo {pages} {148103} (\bibinfo
  {year} {2007})}\BibitemShut {NoStop}%
\bibitem [{\citenamefont {Su\'{a}rez-Vargas}\ \emph {et~al.}(2009)\citenamefont
  {Su\'{a}rez-Vargas}, \citenamefont {Gonz\'{a}lez}, \citenamefont
  {Stefanovska},\ and\ \citenamefont {McClintock}}]{Vargasepl}%
  \BibitemOpen
  \bibfield  {author} {\bibinfo {author} {\bibfnamefont {J.~J.}\ \bibnamefont
  {Su\'{a}rez-Vargas}}, \bibinfo {author} {\bibfnamefont {J.~A.}\ \bibnamefont
  {Gonz\'{a}lez}}, \bibinfo {author} {\bibfnamefont {A.}~\bibnamefont
  {Stefanovska}}, \ and\ \bibinfo {author} {\bibfnamefont {P.~V.~E.}\
  \bibnamefont {McClintock}},\ }\href@noop {} {\bibfield  {journal} {\bibinfo
  {journal} {Euro. Phys. Lett.}\ }\textbf {\bibinfo {volume} {85}},\ \bibinfo
  {pages} {38008} (\bibinfo {year} {2009})}\BibitemShut {NoStop}%
\bibitem [{\citenamefont {Koseska}\ \emph {et~al.}(2010)\citenamefont
  {Koseska}, \citenamefont {Ullner}, \citenamefont {Volkov}, \citenamefont
  {Kurths},\ and\ \citenamefont {Ojalvo}}]{cell}%
  \BibitemOpen
  \bibfield  {author} {\bibinfo {author} {\bibfnamefont {A.}~\bibnamefont
  {Koseska}}, \bibinfo {author} {\bibfnamefont {E.}~\bibnamefont {Ullner}},
  \bibinfo {author} {\bibfnamefont {E.}~\bibnamefont {Volkov}}, \bibinfo
  {author} {\bibfnamefont {J.}~\bibnamefont {Kurths}}, \ and\ \bibinfo {author}
  {\bibfnamefont {J.~G.}\ \bibnamefont {Ojalvo}},\ }\href@noop {} {\bibfield
  {journal} {\bibinfo  {journal} {J. Theoret. Biol.}\ }\textbf {\bibinfo
  {volume} {263}},\ \bibinfo {pages} {189} (\bibinfo {year}
  {2010})}\BibitemShut {NoStop}%
\bibitem [{\citenamefont {Boutle}\ \emph {et~al.}(2007)\citenamefont {Boutle},
  \citenamefont {Taylor},\ and\ \citenamefont {Romer}}]{nino}%
  \BibitemOpen
  \bibfield  {author} {\bibinfo {author} {\bibfnamefont {I.}~\bibnamefont
  {Boutle}}, \bibinfo {author} {\bibfnamefont {R.~H.~S.}\ \bibnamefont
  {Taylor}}, \ and\ \bibinfo {author} {\bibfnamefont {R.~A.}\ \bibnamefont
  {Romer}},\ }\href@noop {} {\bibfield  {journal} {\bibinfo  {journal}
  {American Journal of Physics}\ }\textbf {\bibinfo {volume} {53}},\ \bibinfo
  {pages} {15} (\bibinfo {year} {2007})}\BibitemShut {NoStop}%
\bibitem [{\citenamefont {Lisman}\ and\ \citenamefont {Buzsaki}(2008)}]{brain}%
  \BibitemOpen
  \bibfield  {author} {\bibinfo {author} {\bibfnamefont {J.}~\bibnamefont
  {Lisman}}\ and\ \bibinfo {author} {\bibfnamefont {G.}~\bibnamefont
  {Buzsaki}},\ }\href@noop {} {\bibfield  {journal} {\bibinfo  {journal}
  {Schizophr. Bull.}\ }\textbf {\bibinfo {volume} {34}},\ \bibinfo {pages}
  {974} (\bibinfo {year} {2008})}\BibitemShut {NoStop}%
\bibitem [{\citenamefont {Menck}\ \emph {et~al.}(2014)\citenamefont {Menck},
  \citenamefont {Heitzig}, \citenamefont {Kurths},\ and\ \citenamefont
  {Schellnhuber}}]{kurths_pg}%
  \BibitemOpen
  \bibfield  {author} {\bibinfo {author} {\bibfnamefont {P.~J.}\ \bibnamefont
  {Menck}}, \bibinfo {author} {\bibfnamefont {J.}~\bibnamefont {Heitzig}},
  \bibinfo {author} {\bibfnamefont {J.}~\bibnamefont {Kurths}}, \ and\ \bibinfo
  {author} {\bibfnamefont {H.~J.}\ \bibnamefont {Schellnhuber}},\ }\href@noop
  {} {\bibfield  {journal} {\bibinfo  {journal} {Nat. Commun.}\ }\textbf
  {\bibinfo {volume} {5}},\ \bibinfo {pages} {3969} (\bibinfo {year}
  {2014})}\BibitemShut {NoStop}%
\bibitem [{\citenamefont {Jalife}\ \emph {et~al.}(1998)\citenamefont {Jalife},
  \citenamefont {Gray}, \citenamefont {Morley},\ and\ \citenamefont
  {Davidenko}}]{cardio}%
  \BibitemOpen
  \bibfield  {author} {\bibinfo {author} {\bibfnamefont {J.}~\bibnamefont
  {Jalife}}, \bibinfo {author} {\bibfnamefont {R.~A.}\ \bibnamefont {Gray}},
  \bibinfo {author} {\bibfnamefont {G.~E.}\ \bibnamefont {Morley}}, \ and\
  \bibinfo {author} {\bibfnamefont {J.~M.}\ \bibnamefont {Davidenko}},\
  }\href@noop {} {\bibfield  {journal} {\bibinfo  {journal} {Chaos}\ }\textbf
  {\bibinfo {volume} {8}},\ \bibinfo {pages} {79} (\bibinfo {year}
  {1998})}\BibitemShut {NoStop}%
\bibitem [{\citenamefont {Banerjee}\ and\ \citenamefont
  {Ghosh}(2014{\natexlab{a}})}]{tanpre1}%
  \BibitemOpen
  \bibfield  {author} {\bibinfo {author} {\bibfnamefont {T.}~\bibnamefont
  {Banerjee}}\ and\ \bibinfo {author} {\bibfnamefont {D.}~\bibnamefont
  {Ghosh}},\ }\href@noop {} {\bibfield  {journal} {\bibinfo  {journal} {Phys.
  Rev. E}\ }\textbf {\bibinfo {volume} {89}},\ \bibinfo {pages} {052912}
  (\bibinfo {year} {2014}{\natexlab{a}})}\BibitemShut {NoStop}%
\bibitem [{\citenamefont {Ghosh}\ and\ \citenamefont
  {Banerjee}(2014)}]{tanpre3}%
  \BibitemOpen
  \bibfield  {author} {\bibinfo {author} {\bibfnamefont {D.}~\bibnamefont
  {Ghosh}}\ and\ \bibinfo {author} {\bibfnamefont {T.}~\bibnamefont
  {Banerjee}},\ }\href@noop {} {\bibfield  {journal} {\bibinfo  {journal}
  {Phys. Rev. E}\ }\textbf {\bibinfo {volume} {90}},\ \bibinfo {pages} {062908}
  (\bibinfo {year} {2014})}\BibitemShut {NoStop}%
\bibitem [{\citenamefont {Zakharova}\ \emph {et~al.}(2013)\citenamefont
  {Zakharova}, \citenamefont {Schneider}, \citenamefont {Kyrychko},
  \citenamefont {Blyuss}, \citenamefont {Koseska}, \citenamefont {Fiedler},\
  and\ \citenamefont {Sch{\"{o}}ll}}]{scholl4}%
  \BibitemOpen
  \bibfield  {author} {\bibinfo {author} {\bibfnamefont {A.}~\bibnamefont
  {Zakharova}}, \bibinfo {author} {\bibfnamefont {I.}~\bibnamefont
  {Schneider}}, \bibinfo {author} {\bibfnamefont {Y.~N.}\ \bibnamefont
  {Kyrychko}}, \bibinfo {author} {\bibfnamefont {K.~B.}\ \bibnamefont
  {Blyuss}}, \bibinfo {author} {\bibfnamefont {A.}~\bibnamefont {Koseska}},
  \bibinfo {author} {\bibfnamefont {B.}~\bibnamefont {Fiedler}}, \ and\
  \bibinfo {author} {\bibfnamefont {E.}~\bibnamefont {Sch{\"{o}}ll}},\
  }\href@noop {} {\bibfield  {journal} {\bibinfo  {journal} {Europhysics
  Lett.}\ }\textbf {\bibinfo {volume} {104}},\ \bibinfo {pages} {50004}
  (\bibinfo {year} {2013})}\BibitemShut {NoStop}%
\bibitem [{\citenamefont {Liu}\ \emph {et~al.}(2015)\citenamefont {Liu},
  \citenamefont {Xiao}, \citenamefont {Zhu}, \citenamefont {Zhan},
  \citenamefont {Xiao},\ and\ \citenamefont {Kurths}}]{kurthamp}%
  \BibitemOpen
  \bibfield  {author} {\bibinfo {author} {\bibfnamefont {W.}~\bibnamefont
  {Liu}}, \bibinfo {author} {\bibfnamefont {G.}~\bibnamefont {Xiao}}, \bibinfo
  {author} {\bibfnamefont {Y.}~\bibnamefont {Zhu}}, \bibinfo {author}
  {\bibfnamefont {M.}~\bibnamefont {Zhan}}, \bibinfo {author} {\bibfnamefont
  {J.}~\bibnamefont {Xiao}}, \ and\ \bibinfo {author} {\bibfnamefont
  {J.}~\bibnamefont {Kurths}},\ }\href@noop {} {\bibfield  {journal} {\bibinfo
  {journal} {Phys. Rev. E}\ }\textbf {\bibinfo {volume} {91}},\ \bibinfo
  {pages} {052902} (\bibinfo {year} {2015})}\BibitemShut {NoStop}%
\bibitem [{\citenamefont {Banerjee}\ and\ \citenamefont
  {Ghosh}(2014{\natexlab{b}})}]{tanpre2}%
  \BibitemOpen
  \bibfield  {author} {\bibinfo {author} {\bibfnamefont {T.}~\bibnamefont
  {Banerjee}}\ and\ \bibinfo {author} {\bibfnamefont {D.}~\bibnamefont
  {Ghosh}},\ }\href@noop {} {\bibfield  {journal} {\bibinfo  {journal} {Phys.
  Rev. E}\ }\textbf {\bibinfo {volume} {89}},\ \bibinfo {pages} {062902}
  (\bibinfo {year} {2014}{\natexlab{b}})}\BibitemShut {NoStop}%
\bibitem [{\citenamefont {Majdandzic1}\ \emph {et~al.}(2013)\citenamefont
  {Majdandzic1}, \citenamefont {Podobnik}, \citenamefont {Buldyrev},
  \citenamefont {Kenett}, \citenamefont {Havlin},\ and\ \citenamefont
  {Stanley}}]{ryth1}%
  \BibitemOpen
  \bibfield  {author} {\bibinfo {author} {\bibfnamefont {A.}~\bibnamefont
  {Majdandzic1}}, \bibinfo {author} {\bibfnamefont {B.}~\bibnamefont
  {Podobnik}}, \bibinfo {author} {\bibfnamefont {S.~V.}\ \bibnamefont
  {Buldyrev}}, \bibinfo {author} {\bibfnamefont {D.~Y.}\ \bibnamefont
  {Kenett}}, \bibinfo {author} {\bibfnamefont {S.}~\bibnamefont {Havlin}}, \
  and\ \bibinfo {author} {\bibfnamefont {H.~E.}\ \bibnamefont {Stanley}},\
  }\href@noop {} {\bibfield  {journal} {\bibinfo  {journal} {Nat. Phys.}\
  }\textbf {\bibinfo {volume} {10}},\ \bibinfo {pages} {34} (\bibinfo {year}
  {2013})}\BibitemShut {NoStop}%
\bibitem [{\citenamefont {Morino}\ \emph {et~al.}(2013)\citenamefont {Morino},
  \citenamefont {Tanaka},\ and\ \citenamefont {Aihara}}]{ryth4}%
  \BibitemOpen
  \bibfield  {author} {\bibinfo {author} {\bibfnamefont {K.}~\bibnamefont
  {Morino}}, \bibinfo {author} {\bibfnamefont {G.}~\bibnamefont {Tanaka}}, \
  and\ \bibinfo {author} {\bibfnamefont {K.}~\bibnamefont {Aihara}},\
  }\href@noop {} {\bibfield  {journal} {\bibinfo  {journal} {Phys. Rev. E}\
  }\textbf {\bibinfo {volume} {88}},\ \bibinfo {pages} {032909} (\bibinfo
  {year} {2013})}\BibitemShut {NoStop}%
\bibitem [{\citenamefont {Zou}\ \emph {et~al.}(2015)\citenamefont {Zou},
  \citenamefont {Senthilkumar}, \citenamefont {Nagao}, \citenamefont {Kiss},
  \citenamefont {Tang}, \citenamefont {Koseska}, \citenamefont {Duan},\ and\
  \citenamefont {Kurths}}]{kurthnat15}%
  \BibitemOpen
  \bibfield  {author} {\bibinfo {author} {\bibfnamefont {W.}~\bibnamefont
  {Zou}}, \bibinfo {author} {\bibfnamefont {D.~V.}\ \bibnamefont
  {Senthilkumar}}, \bibinfo {author} {\bibfnamefont {R.}~\bibnamefont {Nagao}},
  \bibinfo {author} {\bibfnamefont {I.~Z.}\ \bibnamefont {Kiss}}, \bibinfo
  {author} {\bibfnamefont {Y.}~\bibnamefont {Tang}}, \bibinfo {author}
  {\bibfnamefont {A.}~\bibnamefont {Koseska}}, \bibinfo {author} {\bibfnamefont
  {J.}~\bibnamefont {Duan}}, \ and\ \bibinfo {author} {\bibfnamefont
  {J.}~\bibnamefont {Kurths}},\ }\href@noop {} {\bibfield  {journal} {\bibinfo
  {journal} {Nat. Commun.}\ }\textbf {\bibinfo {volume} {6}},\ \bibinfo {pages}
  {7709} (\bibinfo {year} {2015})}\BibitemShut {NoStop}%
\bibitem [{\citenamefont {Karnatak}\ \emph {et~al.}(2007)\citenamefont
  {Karnatak}, \citenamefont {Ramaswamy},\ and\ \citenamefont {Prasad}}]{con}%
  \BibitemOpen
  \bibfield  {author} {\bibinfo {author} {\bibfnamefont {R.}~\bibnamefont
  {Karnatak}}, \bibinfo {author} {\bibfnamefont {R.}~\bibnamefont {Ramaswamy}},
  \ and\ \bibinfo {author} {\bibfnamefont {A.}~\bibnamefont {Prasad}},\
  }\href@noop {} {\bibfield  {journal} {\bibinfo  {journal} {Phys. Rev. E}\
  }\textbf {\bibinfo {volume} {76}},\ \bibinfo {pages} {035201R} (\bibinfo
  {year} {2007})}\BibitemShut {NoStop}%
\bibitem [{\citenamefont {Konishi}(2003)}]{dyn}%
  \BibitemOpen
  \bibfield  {author} {\bibinfo {author} {\bibfnamefont {K.}~\bibnamefont
  {Konishi}},\ }\href@noop {} {\bibfield  {journal} {\bibinfo  {journal} {Phys.
  Rev. E}\ }\textbf {\bibinfo {volume} {68}},\ \bibinfo {pages} {067202}
  (\bibinfo {year} {2003})}\BibitemShut {NoStop}%
\bibitem [{\citenamefont {Karnatak}(2015)}]{karnatak_pit}%
  \BibitemOpen
  \bibfield  {author} {\bibinfo {author} {\bibfnamefont {R.}~\bibnamefont
  {Karnatak}},\ }\href@noop {} {\bibfield  {journal} {\bibinfo  {journal}
  {arXiv}\ }\textbf {\bibinfo {volume} {1502.04352[nlin.CD]}} (\bibinfo {year}
  {2015})}\BibitemShut {NoStop}%
\bibitem [{\citenamefont {Banerjee}\ and\ \citenamefont
  {Biswas}(2013)}]{tanchaosad}%
  \BibitemOpen
  \bibfield  {author} {\bibinfo {author} {\bibfnamefont {T.}~\bibnamefont
  {Banerjee}}\ and\ \bibinfo {author} {\bibfnamefont {D.}~\bibnamefont
  {Biswas}},\ }\href@noop {} {\bibfield  {journal} {\bibinfo  {journal}
  {Chaos}\ }\textbf {\bibinfo {volume} {23}},\ \bibinfo {pages} {043101}
  (\bibinfo {year} {2013})}\BibitemShut {NoStop}%
\bibitem [{\citenamefont {Pathria}\ and\ \citenamefont
  {Beale}(2011)}]{pathria}%
  \BibitemOpen
  \bibfield  {author} {\bibinfo {author} {\bibfnamefont {R.~K.}\ \bibnamefont
  {Pathria}}\ and\ \bibinfo {author} {\bibfnamefont {P.~D.}\ \bibnamefont
  {Beale}},\ }\href@noop {} {\emph {\bibinfo {title} {Statistical
  mechanics}}},\ \bibinfo {edition} {3rd}\ ed.\ (\bibinfo  {publisher}
  {Butterworth Heinemann},\ \bibinfo {year} {2011})\BibitemShut {NoStop}%
\bibitem [{\citenamefont {Ermentrout}\ and\ \citenamefont
  {Kopell}(1990)}]{bard}%
  \BibitemOpen
  \bibfield  {author} {\bibinfo {author} {\bibfnamefont {G.~B.}\ \bibnamefont
  {Ermentrout}}\ and\ \bibinfo {author} {\bibfnamefont {N.}~\bibnamefont
  {Kopell}},\ }\href@noop {} {\bibfield  {journal} {\bibinfo  {journal} {SIAM
  J. Appl. Math.}\ }\textbf {\bibinfo {volume} {50}},\ \bibinfo {pages} {125}
  (\bibinfo {year} {1990})}\BibitemShut {NoStop}%
\bibitem [{\citenamefont {Garc{\'{i}}a-Ojalvo}\ \emph
  {et~al.}(2004)\citenamefont {Garc{\'{i}}a-Ojalvo}, \citenamefont {Elowitz},\
  and\ \citenamefont {Strogatz}}]{qstr}%
  \BibitemOpen
  \bibfield  {author} {\bibinfo {author} {\bibfnamefont {J.}~\bibnamefont
  {Garc{\'{i}}a-Ojalvo}}, \bibinfo {author} {\bibfnamefont {M.~B.}\
  \bibnamefont {Elowitz}}, \ and\ \bibinfo {author} {\bibfnamefont {S.~H.}\
  \bibnamefont {Strogatz}},\ }\href@noop {} {\bibfield  {journal} {\bibinfo
  {journal} {Proc. Natl. Acad. Sci. USA}\ }\textbf {\bibinfo {volume} {101}},\
  \bibinfo {pages} {10955} (\bibinfo {year} {2004})}\BibitemShut {NoStop}%
\bibitem [{\citenamefont {Banerjee}\ \emph {et~al.}(2015)\citenamefont
  {Banerjee}, \citenamefont {Dutta},\ and\ \citenamefont {Gupta}}]{bandutta}%
  \BibitemOpen
  \bibfield  {author} {\bibinfo {author} {\bibfnamefont {T.}~\bibnamefont
  {Banerjee}}, \bibinfo {author} {\bibfnamefont {P.~S.}\ \bibnamefont {Dutta}},
  \ and\ \bibinfo {author} {\bibfnamefont {A.}~\bibnamefont {Gupta}},\
  }\href@noop {} {\bibfield  {journal} {\bibinfo  {journal} {Phys. Rev. E}\
  }\textbf {\bibinfo {volume} {91}},\ \bibinfo {pages} {052919} (\bibinfo
  {year} {2015})}\BibitemShut {NoStop}%
\bibitem [{\citenamefont {Ermentrout}(2002)}]{xpp}%
  \BibitemOpen
  \bibfield  {author} {\bibinfo {author} {\bibfnamefont {B.}~\bibnamefont
  {Ermentrout}},\ }\href@noop {} {\emph {\bibinfo {title} {Simulating,
  Analyzing, and Animating Dynamical Systems: A Guide to Xppaut for Researchers
  and Students (Software, Environments, Tools)}}}\ (\bibinfo  {publisher} {SIAM
  Press},\ \bibinfo {year} {2002})\BibitemShut {NoStop}%
\end{thebibliography}
\providecommand{\noopsort}[1]{}\providecommand{\singleletter}[1]{#1}%
\end{document}